\documentclass[oneside,reqno,12pt]{amsart}

\usepackage{latexsym}
\usepackage{rangecite}

\DeclareFontFamily{OT1}{rsfs10}{}
\DeclareFontShape{OT1}{rsfs10}{m}{n}{ <-> rsfs10 }{}
\DeclareMathAlphabet{\mathscript}{OT1}{rsfs10}{m}{n}

\numberwithin{equation}{section}

\newcommand{\du}[3]{{\smash{{#1}_{#2}}^{#3}}}
\newcommand{\ud}[3]{{\smash{{#1}^{#2}}_{#3}}}
\newcommand{\sud}[3]{\smash{#1^{#2}_{#3}}}

\newcommand{\SL}{\mathrm{SL}}
\newcommand{\SU}{\mathrm{SU}}

\newcommand{\cc}{s}

\makeatletter
    \def\serieslogo@{\vtop to 
0pt{\noindent\scriptsize\ppn\parindent\z@}}
    \let\@setcopyright\@empty
\makeatother

\begin{document}

\def\ppn{HEP-TH/9609016, UPR-712T}

\title[Vector-Tensor multiplet in $N=2$ superspace with central 
charge]{Vector-Tensor multiplet in $N=2$ superspace\\ with central 
charge}
\author{Ahmed Hindawi, Burt A. Ovrut, and Daniel Waldram}
\maketitle
\vspace*{-0.3in}
\begin{center}
\small{\textit{Department of Physics, University of Pennsylvania}} \\
\small{\textit{Philadelphia, PA 19104-6396, USA}}
\end{center}

\begin{abstract}

We use the four-dimensional $N=2$ central charge superspace to give a 
geometrical construction of the Abelian vector-tensor multiplet 
consisting, under $N=1$ supersymmetry, of one vector and one linear 
multiplet. We derive the component field supersymmetry and central 
charge transformations, and show that there is a super-Lagrangian, the 
higher components of which are all total derivatives, allowing us to 
construct superfield and component actions.

\vspace*{\baselineskip}


 
\end{abstract}

\renewcommand{\baselinestretch}{1.2} \large{} \normalsize{}

\vspace*{\baselineskip}

\section{Introduction}

It has long been known that there is an $N=2$ supermultiplet which, 
with respect to $N=1$ supersymmetry, appears as a combination of one 
vector and one chiral multiplet 
\cite{NPB-79-413,NPB-113-135,NPB-121-77,NPB-133-275}. This 
supermultiplet is irreducible under $N=2$ supersymmetry and carries 
vanishing central charge. It is usually referred to as the vector 
supermultiplet. Not long after its introduction, it was realized that 
there exists a variant, the vector-tensor multiplet, which, under $N=1$ 
supersymmetry, appears as a combination of one vector and one linear 
multiplet \cite{PLB-92-123,NPB-173-127}. This variant is irreducible 
under $N=2$ and, in contrast, carries non-vanishing central charge 
off-shell. On shell it is equivalent to the vector multiplet. This 
multiplet was originally constructed using component field techniques 
\cite{PLB-92-123,NPB-173-127}, and it and its properties have received 
little attention. Recently, however, within the context of trying to 
understand the consequences of $N=2$ duality in superstrings, the 
vector-tensor multiplet has re-emerged. Specifically, when heterotic 
string theory is reduced to an $N=2$ theory in four dimensions the 
dilaton and the antisymmetric tensor field lie in a vector-tensor 
multiplet \cite{NPB-451-53}. It is clear from this work that the 
vector-tensor supermultiplet is fundamental in heterotic theories and 
that elucidation of its properties, such as its behavior under duality 
transformations, is of importance. The coupling of the vector-tensor 
multiplet to supergravity has also recently been considered 
\cite{PLB-373-81}. It was shown that this requires the gauging of the 
central charge, leading to a Chern-Simons coupling between the 
vector-tensor multiplet and a vector multiplet.

If one wants to fully understand the complete, off-shell structure of 
the tensor-vector multiplet, one powerful approach is to construct the 
appropriate $N=2$ superfield. It is the purpose of this paper to 
provide such a formulation. Much of the interesting structure of the 
vector-tensor multiplet appears in its couplings to other multiplets 
\cite{PLB-373-81}. Here, however, we will restrict ourselves to a 
superfield formulation of the free Abelian case, leaving interacting 
generalization to future publications. Since the vector-tensor 
supermultiplet has non-zero central charge, it is necessary to expand 
the usual superspace. The maximal central extension of the $N=2$ 
superalgebra has two central charges. Consequently the corresponding 
superspace has two extra bosonic coordinates \cite{NPB-138-109}. 
Working in this superspace, one can introduce superfields and covariant 
constraints. Following the geometrical formulation of supergauge 
fields, we introduce a super-connection and the associated curvature. 
We then introduce the appropriate constraints, solve the Bianchi 
identities and show how the component fields of the tensor-vector 
multiplet emerge. Using superspace techniques we rederive the 
supersymmetry and central charge transformations and, after showing 
that there exist a Lagrangian superfield in central charge superspace, 
the higher components of which are all total derivatives, we give the 
superfield and component actions. The existence of a superfield 
Lagrangian is an example of a central-charge generalization of the 
superactions described in \cite{NPB-191-445}. It suggests that there is 
an even-dimensional submanifold of the central charge superspace
naturally associated with the vector-tensor multiplet. 

Central charge superspace has recently also been considered in 
\cite{PLB-373-89,Gaida96b} and used to rederive the usual gauge 
supermultiplet. Also, we want to point out that there is a strong
relationship between the theory of $N=1$ supersymmetry in six 
dimensions and four-dimensional, $N=2$ central charge superspace. 
Indeed, the superfield equations of motion in the six-dimensional 
theory motivated the choice of one important constraint equation in 
four-dimensional central charge superspace. In this sense, our 
constraint can be understood as a superfield realization of the 
dimensional reduction by Legendre transformation discussed in 
\cite{NPB-173-127}.

\section{$N=2$ Superspace}

We start by briefly recapitulating the relevant formulae of $N=2$ 
superspace. Throughout the paper we will use the conventions of 
\cite{NPB-138-109,NPB-133-275,WB-SS}. The $N=2$ supersymmetry algebra 
is obtained from the Poincar\'e algebra by adding four fermionic 
operators $Q_\alpha^i$ ($\alpha=1,2$ and $i=1,2$) and their 
anti-hermitian conjugates $\bar Q_{\dot\alpha i} = - 
(Q^i_\alpha)^\dag$. Moreover the algebra admits one complex ``central 
charge'' $Z = Z_1 + i Z_2$ where $Z_1$ and $Z_2$ are hermitian. These 
operators satisfy the following anticommutation relations:
\begin{equation}
\begin{split}
\{ Q_\alpha^i , \bar Q_{\dot\alpha j} \} &= 2 \ud\delta i j 
\ud\sigma m {\alpha\dot\alpha} P_m, \\
\{ Q_\alpha^i , Q_\beta^j \} &= 2 \epsilon_{\alpha\beta} \epsilon^{ij} 
Z, \\
\{ \bar Q_{\dot\alpha i} , \bar Q_{\dot\beta j} \} &= -2 
\epsilon_{\dot\alpha 
\dot\beta} \epsilon_{ij} \bar Z,
\end{split}
\label{Qalg}
\end{equation}
where $P_m$ are the four-momenta and $\bar Z = Z^\dag$. The 
automorphism group of algebra \eqref{Qalg} is $\SL(2,\mathbb C) \otimes 
\SU(2)$ where the $\SU(2)$ acts on the $i$ index in $Q_\alpha^i$. The 
antisymmetric tensor $\epsilon_{ij}$ with $\epsilon_{12}=-1$ provides 
an invariant metric for raising and lowering the $\SU(2)$ indices by 
$a_i=\epsilon_{ij}a^j$ and $a^i=\epsilon^{ij}a_j$. 

$N=2$ superspace is a space with coordinates $z^M = 
(x^m,\theta^\alpha_i,\bar\theta_{\dot\alpha}^i,z,\bar z)$ where $x^m$, 
$z$, and $\bar z=z^*$ are commuting bosonic coordinates while 
$\theta^\alpha_i$ and $\bar\theta_{\dot\alpha}^i=(\theta_{\alpha i})^*$ 
are anticommuting fermionic coordinates. A superfield $\phi$ is a 
function of the superspace coordinates, $\phi = \phi(x^m, 
\theta^\alpha_i, \bar\theta_{\dot\alpha}^i, z, \bar z)$. Taylor-series 
expansion of a general superfield $\phi$ in the $\theta$ coordinates 
terminates after a finite number of terms due to the anticommuting 
nature of $\theta$. On the other hand, the expansion in $z$ and $\bar 
z$ never ends. This means there are an infinite number of component 
fields (functions of the spacetime coordinates $x^m$) in a general 
superfield. This infinite number of component fields can be reduced to 
a finite number either off-shell, by applying appropriate constraints, 
or on-shell, by choosing equations of motion which propagate only a 
finite number of component fields.

Translations in the superspace are generated by the supercovariant 
differential operators $\partial_a$, $\partial_z$, $\partial_{\bar z}$, 
and
\begin{equation}
\begin{split}
D_\alpha^i &= \frac{\partial}{\partial \theta^\alpha_i} + i \ud\sigma a 
{\alpha\dot\alpha} \bar\theta^{\dot\alpha i} \partial_a - i 
\theta_\alpha^i \partial_z, \\
\bar D_{\dot\alpha i} &= - \frac{\partial}{\partial 
\bar\theta^{\dot\alpha i}} - i \theta^\alpha_i \ud\sigma a 
{\alpha\dot\alpha} \partial_a - i \bar\theta_{\dot\alpha i} 
\partial_{\bar z},
\end{split}
\end{equation}
where $\bar D_{\dot\alpha i}=-(D^i_\alpha)^\dag$. It is straightforward 
to compute the anticommutation relations for these operators. They are
\begin{equation}
\begin{split}
\{ D_\alpha^i , \bar D_{\dot\alpha j} \} &= - 2i \ud\delta i j \ud 
\sigma m {\alpha\dot\alpha} \partial_m, \\
\{ D_\alpha^i , D_\beta^j \} &= -2i \epsilon_{\alpha\beta} 
\epsilon^{ij} \partial_z, \\
\{ \bar D_{\dot\alpha i} , \bar D_{\dot\beta j} \} &= 2i 
\epsilon_{\dot\alpha\dot\beta} \epsilon_{ij} \partial_{\bar z}.
\end{split}
\label{Dalg}
\end{equation}
By construction $D^i_\alpha$ and $\bar D_{\dot\alpha i}$ anticommute 
with $Q^i_\alpha$ and $\bar Q_{\dot\alpha i}$ and so can be used to 
impose supersymmetric covariant conditions on superfields.

The supervielbein $\du e A M$ of the $N=2$ superspace is defined as the 
matrix that relates the supercovariant derivatives $D_A=(\partial_a, 
D^\alpha_i, \bar D_{\dot \alpha i}, \partial_z, \partial_{\bar z})$ and 
the ordinary partial derivatives
\begin{equation}
D_A = \du e A M \frac\partial{\partial z^M}.
\end{equation}
The matrix $\du e M A$ is the inverse of $\du e A M$. These two 
matrices define the geometry of the $N=2$ superspace. The torsion $T^A$ 
is defined as the exterior derivative of the supervielbein one-form 
$e^A=dz^M \du eMA$,
\begin{equation}
T^A = d e^A  = \tfrac12 e^C e^B \du T {BC} A.
\end{equation}
The non-vanishing components of the torsion are found to be
\begin{equation}
\begin{split}
\smash{\smash{T^i_\alpha}^j_{\dot\alpha}}^a &= {\sud{\sud T j 
{{\dot\alpha}}} i \alpha}^a = - 2 i \epsilon^{ij} \ud\sigma a 
{\alpha\dot\alpha}, \\
\sud{\sud T i \alpha}j\beta^z  &= \sud{\sud T j\beta} i \alpha^z = 2 i  
\epsilon^{ij} \epsilon_{\alpha\beta}, \\
\sud{\sud T i {\dot\alpha}} j {\dot\beta}^{\bar z} &= \sud{\sud T j 
{\dot\beta}} i {\dot\alpha}^{\bar z} =  2 i \epsilon^{ij} 
\epsilon_{\dot\alpha\dot\beta}. 
\end{split}
\end{equation}

\section{Vector-Tensor Multiplet in $N=2$ Superspace}

In this section we give a geometrical superfield formulation of the 
vector-tensor multiplet. We begin by considering the usual geometrical 
form of super-gauge theory, though in superspace with central charge. 
We then find a suitable set of constraints on the superfield strength 
to reproduce the field content of the vector-tensor multiplet. 
Restricting our attention to Abelian gauge theories, let us introduce a 
hermitian connection $A=dz^M A_M = e^A A_A$. The hermiticity of the 
connection implies
\begin{equation}
A = dx^a A_a + d\theta^\alpha_i A^i_\alpha + d\bar\theta^i_{\dot\alpha} 
\bar A_i^{\dot\alpha} + dz A_z + d\bar z A_{\bar z},
\end{equation}
where $A_a$ is real, $\bar A^{\dot\alpha}_i = (A^{\alpha i})^\dag$, and 
$A_{\bar z} = A_z^\dag$. The curvature two-form is defined as
\begin{equation}
F = dA = \tfrac12 e^B e^A F_{AB}.
\end{equation}
The coefficient functions $F_{BA}$ comprise thirteen Lorentz-covariant 
types. The ones that contain torsion terms are given by
\begin{equation}
\begin{split}
\sud{\sud F i \alpha} j {\dot\alpha} &= D_\alpha^i A_{\dot\alpha}^j + 
\bar D_{\dot\alpha}^j \bar A_\alpha^i - 2i \epsilon^{ij} \ud \sigma a 
{\alpha\dot\alpha} A_a, \\
\sud{\sud F i \alpha} j \beta &= D_\alpha^i A_\beta^j + D_\beta^i 
A_\alpha^j + 2 i \epsilon_{\alpha\beta} \epsilon^{ij} A_z, \\
\sud{\sud F i {\dot\alpha}} j {\dot\beta} &= \bar D_{\dot\alpha}^i \bar 
A_{\dot\beta}^j + \bar D_{\dot\beta}^i \bar A_{\dot\alpha}^j + 2 i 
\epsilon_{\dot\alpha\dot\beta} \epsilon^{ij} A_{\bar z},
\end{split}
\end{equation}
The curvature tensor $F$ is subject to the Bianchi identities $dF=0$ or 
equivalently 
\begin{equation}
D_{A}F_{BC} + \du T {AB} D F_{DC} + \text{graded cycle} = 0.
\end{equation}

Each tensor component of the curvature tensor is an $N=2$ superfield 
which, in turn, has an infinite number of component fields. All but a 
finite number of these component fields must be eliminated by virtue of 
appropriate constraints. These constraints must be Lorentz, gauge, and 
supersymmetric covariant. It is natural to adopt a set of constraints 
which set the pure spinorial part of the curvature tensor to zero. That 
is 
\begin{equation}
\sud{\sud F i \alpha} j \beta = \sud{\sud F i \alpha} j {\dot\beta} = 
\sud{\sud F i {\dot\alpha}} j {\dot\beta} = 0.
\label{con-1}
\end{equation}
Before imposing further conditions we would like to explore the 
consequences of \eqref{con-1}. To do so we must solve the Bianchi 
identities subject to these constraints. The result is that all the 
components of the curvature tensor $F_{AB}$ are determined in terms of 
a single superfield $F^i_{\alpha\bar z}$ and its hermitian conjugate 
$F_{\dot\alpha i z} = (F^i_{\alpha \bar z})^\dag$. Henceforth, we 
denote these superfields by $W^i_\alpha$ and $\bar W_{\dot\alpha i}$. 
In particular $F^i_{\alpha z} = 0$ and
\begin{align} 
F_{ab} &= -\tfrac1{16} i \epsilon^{ij} 
\du{\bar\sigma}a{\dot\alpha\alpha} \du{\bar\sigma}b{\dot\beta\beta} 
\left( \epsilon_{\dot\alpha\dot\beta} D^j_\beta W^i_\alpha + 
\epsilon_{\alpha\beta} \bar D^j_{\dot\beta} \bar W^i_{\dot\alpha} 
\right). \label{3.32} \\
F^i_{\dot\alpha a} &= - \tfrac12 \epsilon_{\dot\alpha\dot\beta} 
\du{\bar\sigma}a{\dot\beta\alpha} W^i_\alpha, \label{3.31} \\ 
F_{a\bar z} &= - \tfrac18 i \epsilon_{ij} 
\du{\bar\sigma}a{\dot\beta\alpha} \bar D^j_{\dot\beta} W^i_\alpha, 
\label{3.30} \\
F_{z\bar z}  &= \tfrac14 i \epsilon_{ij} \epsilon^{\alpha\beta} 
D_\alpha^i W_\beta^j, \label{3.29}
\end{align}
Furthermore, $W^i_\alpha$ is constrained to satisfy 
\begin{align}
& D^{(j}_{(\beta} W^{i)}_{\alpha)} = 0, \qquad \bar D^{(j}_{(\dot\beta} 
\bar W^{i)}_{\dot\alpha)} = 0, \label{1} \\
& \bar D^{(j}_{\dot\beta} W^{i)}_{\alpha} = 0, \qquad D^{(j}_\beta \bar 
W^{i)}_{\dot\alpha} = 0, \label{2} \\
& \epsilon^{\alpha\beta} D_\alpha^{[i} W_\beta^{j]} 
= - \epsilon^{\dot\alpha\dot\beta} \bar D_{\dot\alpha}^{[i} \bar 
W_{\dot\beta}^{j]}, \label{3} \\
& \epsilon^{\alpha\beta} D_\alpha^{(i} W_\beta^{j)} 
= \epsilon^{\dot\alpha\dot\beta} \bar D_{\dot\alpha}^{(i} \bar 
W_{\dot\beta}^{j)}.
\label{4}
\end{align}

Let us explore the consequences of \eqref{1}--\eqref{4} on the field 
content of $W^i_\alpha$. The expansion of $W^i_\alpha$ in the 
anti-commuting coordinates has the general form
\begin{equation}
W^i_\alpha = \lambda^i_\alpha + \theta_j^\beta \sud{\sud G i \alpha} j 
\beta + \bar\theta_j^{\dot\alpha} \sud{\sud H i \alpha} j {\dot\alpha} 
+ \mathcal{O}(\theta^2).
\end{equation}
It is obvious that conditions \eqref{1}--\eqref{4} will not impose any 
restriction on $\lambda^i_\alpha$. First we consider the implications 
of the lowest component of superfield constraints \eqref{1}--\eqref{4}. 
Condition \eqref{1} implies that
\begin{equation}
\sud{\sud G i \alpha} j \beta = i \epsilon^{ij} f_{\alpha\beta} 
+ 2\epsilon^{ij}\epsilon_{\alpha\beta} D + i \epsilon_{\alpha\beta} 
\rho^{ij}, 
\end{equation}
where $f_{\alpha\beta}=f_{(\alpha\beta)}$ and $\rho^{ij}=\rho^{(ij)}$. 
Conditions \eqref{3} and \eqref{4} further implies the reality 
condition $D=D^\dag$ and $\rho^{ij}=\bar\rho^{ij}$ where 
$\bar\rho_{ij}=(\rho^{ij})^\dag$. Condition \eqref{2} yields
\begin{equation}
\sud{\sud H i \alpha} j {\dot\alpha} = i \epsilon^{ij} 
h_{\alpha\dot\alpha}.
\end{equation}

Higher components of the superfield constraint \eqref{1}--\eqref{4} 
imply further conditions on the fields $f_{\alpha\beta}$, 
$h_{\alpha\dot\alpha}$, and $D$. One way to realize these conditions is 
to note that, from equations \eqref{3.32}, \eqref{3.30}, and 
\eqref{3.29},
\begin{align}
\mathscript{F}_{ab} &= F_{ab}| = - \tfrac18 \du{\bar\sigma} a 
{\dot\alpha\alpha} \du{\bar\sigma} b {\dot\beta\beta} \left( 
\epsilon_{\dot\alpha\dot\beta} f_{\alpha\beta} + \epsilon_{\alpha\beta} 
\bar f_{\dot\alpha\dot\beta} \right), \\
\left. F_{a\bar z} \right| &= \tfrac12 h_a, \label{F1} \\
F_{z\bar z}| &= 2 i D, \label{F2}
\end{align}
where $\bar f_{\dot\alpha\dot\beta} = (f_{\alpha\beta})^\dag$ and $h_a 
= -\tfrac12 \du \sigma a {\alpha\dot\alpha} h_{\alpha\dot\alpha}$. The 
equations \eqref{3.29}--\eqref{4} are the general solution of the 
Bianchi identities subject to our constraints. Three of the Bianchi 
identities they satisfy are
\begin{align}
\partial_a F_{bc} + \partial_b F_{ca} + \partial_c F_{ab} &=0, 
\label{ebi1} \\
\partial_a F_{b\bar z} + \partial_b F_{\bar z a} + \partial_{\bar z} 
F_{ab} &= 0, \label{ebi2} \\
\partial_a F_{z\bar z} + \partial_z F_{\bar z a} + \partial_{\bar z} 
F_{az} &= 0. \label{ebi3}
\end{align}
Taking the lowest component of equation \eqref{ebi1} gives rise to
\begin{equation}
\partial_a \mathscript F_{bc} + \partial_b \mathscript F_{ca} + 
\partial_c \mathscript F_{ab} = 0,
\end{equation}
so that $\mathscript F_{ab}$ is the field strength of a gauge field 
$V_a$. That is 
\begin{equation}
\mathscript F_{ab}=\partial_a V_b - \partial_b V_a.
\end{equation}
The lowest components of \eqref{ebi2} and \eqref{ebi3} give
\begin{align}
\partial_{[a} h^R_{b]} &= - \tfrac12 (\partial_z + \partial_{\bar z}) 
\mathscript F_{ab}, \\
\partial_{[a} h^I_{b]} &= - \tfrac12 i (\partial_z-\partial_{\bar z}) 
\mathscript F_{ab}, \label{h1} \\
\partial_a D &= - \tfrac14 i \left\{ (\partial_z-\partial_{\bar z}) 
h_a^R  + i (\partial_z + \partial_{\bar z}) h_a^I \right\},
\label{temp11}
\end{align}
where $h_a = h_a^R + i h_a^I$.

This is as far as we can go using solely the constraints \eqref{con-1}. 
So far we have the following component fields: an $\SU(2)$ doublet of 
spinors $\lambda^i_{\alpha}$, a real gauge field $V_a$, a complex 
vector field $h_a$, a real scalar $D$ and a real $\SU(2)$ triplet of 
scalars $\rho^{ij}$. We would like to impose further constraints to 
reduce the number of fields to an irreducible multiplet. As we will 
show very shortly, it is possible to impose further constraints to 
ensure that the real and imaginary parts of $h_a$ are each field 
strengths, one of an anti-symmetric tensor and the other of a scalar. 
What about the fields $D$ and $\rho^{ij}$? We will find that $D$ and 
$\rho^{ij}$ play the role of auxiliary fields. Note that the usual 
$N=2$ gauge multiplet has a triplet of auxiliary fields. Hence, if we 
want to derive the usual gauge multiplet we are led to set $D$ to zero 
by promoting the constraint \eqref{3} to the stronger one
\begin{equation}
\epsilon^{\alpha\beta} D_\alpha^{[i} W_\beta^{j]} = 0, \qquad  
\epsilon^{\dot\alpha\dot\beta} \bar D_{\dot\alpha}^{[i} \bar 
W_{\dot\beta}^{j]} = 0.
\end{equation}
On the other hand, if we want to derive the vector-tensor multiplet, 
which has a single auxiliary field, we must eliminate $\rho^{ij}$. This 
can be achieved by promoting condition \eqref{4} to the stronger 
condition
\begin{equation}
\epsilon^{\alpha\beta} D_\alpha^{(i} W_\beta^{j)} = 0, \qquad 
\epsilon^{\dot\alpha\dot\beta} \bar D_{\dot\alpha}^{(i} \bar 
W_{\dot\beta}^{j)} = 0.
\label{4'}
\end{equation}
Henceforth, we add condition \eqref{4'} as a further constraint on the 
theory. Now let us proceed using conditions \eqref{1}--\eqref{3} as 
well as the stronger version of \eqref{4}, constraint \eqref{4'}. Apart 
from eliminating $\rho^{ij}$, it can be shown that the consequences of 
imposing \eqref{4'} are
\begin{equation}
\begin{split}
\partial^a h_a^R &= i (\partial_z - \partial_{\bar z}) D, \\
\partial^a h_a^I &= - (\partial_z + \partial_{\bar z}) D, \\
\partial^b \mathscript F_{ab} &= -\tfrac14 \left\{ 
(\partial_z+\partial_{\bar z}) h_a^R + i (\partial_z - \partial_{\bar 
z}) h_a^I \right\}.
\end{split}
\label{h2}
\end{equation}
We note that the conditions obtained so far on $h_a$ given in 
\eqref{h1} and \eqref{h2} are not sufficient to render the real and 
imaginary components of $h_a$ as field strengths. We would like, 
therefore, to impose yet another, and final, constraint that will 
achieve this goal. We take a reality condition on the central charge
\begin{equation}
\partial_z W^i_\alpha = \partial_{\bar z} W^i_\alpha.
\label{realz}
\end{equation}
This means that superfield $W^i_\alpha$, and hence all the curvature 
tensor components, depend on the two central charges $z$ and $\bar z$ 
only through the combination $z + \bar z$ with no dependence on $z-\bar 
z$. Substituting from constraint \eqref{realz} into \eqref{h1} and 
\eqref{h2} immediately gives the following constraints on the real and 
imaginary components of $h_a$
\begin{equation}
\partial^a h_a^R = 0, \qquad \partial_{[a} h_{b]}^I = 0.
\label{t5}
\end{equation}
These equations achieve just the goal we were aiming for. They assert 
that $h_a^R$ and $h_a^I$ are field strengths of an anti-symmetric 
tensor and a scalar field respectively. That is
\begin{equation}
\begin{split}
h_a^R &= \tfrac13 \epsilon_{abcd} H^{bcd} = \tfrac13 \epsilon_{abcd} 
\partial^b B^{cd}, \\
h_a^I &= 2 \partial_a \phi.
\end{split}
\end{equation}

In conclusion, we find that the component fields in $W^i_\alpha$ are 
exactly those of the vector-tensor multiplet, namely 
$(\lambda^i_\alpha, \phi, B_{ab}, V_a, D)$. It will not have escaped 
the readers notice that these fields are actually functions of the 
central charge coordinates as well as the spacetime coordinates. 
However, by virtue of constraint \eqref{realz}, they depend on the 
central charge coordinates only via $z+\bar z$. Let us introduce a real 
central charge coordinate $\cc=z+\bar z$. Moreover, the dependence of 
these fields on $\cc$ is highly restricted by the conditions we derived 
so far. In fact, as we now show, these conditions completely determine 
their dependence on $\cc$ in terms of their lowest component in their 
central charge expansion. To see this let us summarize the relevant 
conditions in \eqref{h1}, \eqref{temp11}, and \eqref{h2} derived on the 
bosonic fields. They are
\begin{equation}
\begin{split}
\partial_\cc D &= - \tfrac12 \partial^a h_a^I, \\
\partial_\cc h_a^R &= -2 \partial^b \mathscript F_{ab}, \\
\partial_\cc h_a^I &= 2 \partial_a D, \\
\partial_\cc \mathscript F_{ab} &= - \partial_{[a} h_{b]}^R.
\end{split}
\label{t4}
\end{equation}
Equations \eqref{t4} constitute a system of first-order ``differential 
equations'' for the $s$ dependence of the bosonic fields. The general 
solution is completely determined in terms of the ``initial 
conditions''; that is, the values of the fields at $s=0$, say, 
$D(x^m)$, $h_a(x^m)$, and $\mathscript F_{ab}(x^m)$. Thus in a Taylor 
series expansion in $\cc$ each term in the series gets related to 
lower-order terms, so that for instance,
\begin{equation}
D(x^m,\cc) = D(x^m) - \tfrac12 \partial^a h_a^I(x^m) \cc - \tfrac12 
\Box D(x^m) \cc^2 + \cdots,
\end{equation}
with similar formulas for $h_a^R$, $h_a^I$, and $\mathscript F_{ab}$. 
Furthermore, in the next section, we will derive the corresponding 
equation for the fermionic field $\lambda^i_\alpha$.

\section{Supersymmetry and Central Charge Transformations}

In this section, we would like to derive the supersymmetry and central 
charge transformations of the component fields. In order to exhibit 
these transformations it is necessary to compute the part of 
$W^i_\alpha$ quadratic in the fermionic coordinates. Moreover, since 
all the component fields are present in the lowest and first order 
terms of $W^i_\alpha$, the quadratic term is sufficient for computing 
the supersymmetry transformations. We have
\begin{equation}
W^i_\alpha = \lambda^i_\alpha + i \theta^{\beta i}f_{\alpha\beta} + 2 
\theta^i_\alpha D + i \bar\theta^{\dot\alpha i} h_{\alpha\dot\alpha} + 
\theta^\beta_j \theta^\gamma_k \sud{\sud{\sud \Lambda i \alpha} j 
\beta} k \gamma + \theta^\beta_j \bar\theta^{\dot\gamma}_k 
\sud{\sud{\sud \Pi i \alpha} j \beta} k {\dot\gamma} + 
\bar\theta^{\dot\beta}_j \bar\theta^{\dot\gamma}_k \sud{\sud{\sud 
\Sigma i \alpha} j {\dot\beta}} k {\dot\gamma} + \mathcal{O}(\theta^3)
\end{equation}
Let us first consider the constraint equations \eqref{1}, \eqref{2}, 
and  \eqref{4'}. Demanding that the linear term in $\theta$ in these 
equations vanish completely determines $\Lambda$, $\Pi$, and $\Sigma$. 
We find
\begin{equation}
\begin{split}
\sud{\sud{\sud \Lambda i \alpha} j \beta} k \gamma &= \tfrac12 i 
\epsilon_{\beta\gamma} \epsilon^{ij} \partial_\cc \lambda^k_\alpha + 
\tfrac12 i \epsilon_{\beta\gamma} \epsilon^{ik} \partial_\cc 
\lambda^j_\alpha, \\
\sud{\sud{\sud \Pi i \alpha} j \beta} k {\dot\gamma} &= \tfrac12 i 
\epsilon^{jk} \ud\sigma a {\beta\dot\gamma} \partial_a \lambda^i_\alpha 
- \tfrac12 i \epsilon^{ij} \ud\sigma a {\beta\dot\gamma} \partial_a 
\lambda^k_\alpha - \tfrac32 i \epsilon^{ik} \ud\sigma a 
{\beta\dot\gamma} \partial_a \lambda^j_\alpha, \\
\sud{\sud{\sud \Sigma i \alpha} j {\dot\beta}} k {\dot\gamma} &= 
\tfrac12 i \epsilon_{\dot\beta\dot\gamma} \epsilon^{ij} \partial_\cc 
\lambda^k_\alpha + \tfrac12 i \epsilon_{\dot\beta\dot\gamma} 
\epsilon^{ik} \partial_\cc \lambda^j_\alpha.
\end{split}
\label{quad}
\end{equation}
We note that these terms do not involve any new fields. They are 
completely given in terms of the lowest component of $W^i_\alpha$, 
namely $\lambda^i_\alpha$. We are still left with equation \eqref{3} 
which is a reality condition. Evaluating the linear term in \eqref{3}, 
and making use of \eqref{quad}, yields the following constraint on the 
fermionic field $\lambda^i_\alpha$
\begin{equation}
\partial_\cc \lambda^i_\alpha = \ud\sigma a {\alpha\dot\alpha} 
\partial_a \bar\lambda^{\dot\alpha i},
\label{ccl}
\end{equation}
where $\bar\lambda_{\dot\alpha i} = (\lambda^i_{\alpha})^\dag$. 
Equation \eqref{ccl} is the relation for the fermionic field 
$\lambda_i^\alpha$ corresponding to equations \eqref{t4} for the 
bosonic fields. It fixes the expansion of $\lambda^i_\alpha$ in the 
central charge $\cc$, leaving only the lowest component 
$\lambda^i_\alpha(x^m)$ arbitrary.

We are now in a position to compute the supersymmetry transformation of 
the different component fields. To do this we have to act on 
$W^i_\alpha$ with
\begin{equation}
\delta_\xi = \xi^\alpha_i Q_\alpha^i + \bar\xi_{\dot\alpha}^i \bar 
Q_i^{\dot\alpha}.
\end{equation}
We find
\begin{equation}
\begin{split}
\delta_\xi D & = - \tfrac12 i \left( \xi^\alpha_i \ud\sigma a 
{\alpha\dot\alpha} \partial_a \bar\lambda^{\dot\alpha i} + 
\bar\xi^{\dot\alpha}_i \ud\sigma a {\alpha\dot\alpha} \partial_a 
\lambda^{\alpha i} \right), \\
\delta_\xi \phi &= \tfrac12 i \left( \xi_{\alpha i} \lambda^{\alpha i} 
- \bar\xi_{\dot\alpha i} \bar\lambda^{\dot\alpha i} \right), \\
\delta_\xi B_{cd} &= \tfrac16 \epsilon_{abcd} \ud \sigma {ab} 
{(\alpha\beta)} \xi^\alpha_i \lambda^{\beta j} + \tfrac16 
\epsilon_{abcd} \ud {\bar\sigma} {ba} {(\dot\alpha\dot\beta)} 
\bar\xi^{\dot\alpha}_i \bar\lambda^{\dot\beta i}, \\
\delta_\xi V_a & = - \tfrac12 \xi_{\alpha i} \du {\bar\sigma} a 
{\dot\alpha\alpha} \bar \lambda^k_{\dot\alpha} + \tfrac12 
\bar\xi_{\dot\alpha i} \du{\bar\sigma} a {\dot\alpha\alpha} 
\lambda^k_\alpha, \\
\delta_\xi \lambda^i_\alpha & = 2 i \xi^{\beta i} 
\epsilon^{\dot\alpha\dot\beta} \ud\sigma a {\alpha\dot\alpha} \ud\sigma 
b {\beta\dot\beta} \partial_{[a} V_{b]} 
+ 2 \xi_\alpha^i D 
+ \tfrac{i}{3} \bar\xi^{\dot\alpha i} \ud\sigma a {\alpha\dot\alpha} 
\epsilon_{abcd} \partial^b B^{cd} - 2 \ud \sigma a {\alpha\dot\alpha} 
\bar\xi^{\dot\alpha} \partial_a \phi.
\end{split}
\label{susy}
\end{equation}
Note that the fields in these supersymmetry transformations are 
formally functions of both $x^m$ and $\cc$. We have already shown that 
the only independent components of these fields are the lowest order 
terms in $\cc$. Further, transformations \eqref{susy} do not mix 
component fields from different orders in $\cc$. Thus equations 
\eqref{susy} also express the variations of the lowest-order 
independent fields, which are functions of $x^m$ only.

To get the central charge transformations we have to act with 
$\delta_\omega= \omega\partial_\cc$ on $W^i_\alpha$. It is 
straightforward to get the following set of transformations
\begin{equation}
\begin{split}
\delta_\omega D & = - \omega \Box \phi, \\
\delta_\omega \phi &= \omega D, \\
\delta_\omega B^{cd} &= 3 \omega \epsilon^{abcd} \partial_{a} V_b,\\
\delta_\omega V_a & = - \tfrac{1}{6} \omega \epsilon_{abcd} \partial^b 
B^{cd}, \\ 
\delta_\omega \lambda^i_\alpha & = \omega \ud\sigma a 
{\alpha\dot\alpha} \partial_a \bar\lambda^{\dot\alpha i}.
\end{split}
\label{cc}
\end{equation}

Expressions \eqref{susy} and \eqref{cc} reproduce the supersymmetry and 
central charge transformations of the vector-tensor multiplet given in 
the component field calculations of 
\cite{PLB-92-123,NPB-173-127,NPB-451-53}. They are a realization of the 
$N=2$ supersymmetry algebra with central charge. Note that as a result 
of our constraints, in particular \eqref{realz}, only one of the two 
central charges is represented nontrivially. 

We now discuss how to write a superfield action for the vector-tensor 
multiplet. We start by noting that the relevant field strengths of the 
different bosonic fields occur at the linear level in $W^i_\alpha$. To 
get a quadratic expression in these field strengths, we have to 
consider expressions quadratic in $W^i_\alpha$. These expressions can 
contain both $D^i_\alpha$ and $\bar D_{\dot\alpha i}$. Contracting the 
indices to get a proper Lorentz invariant Lagrangian implies that there 
must be an even number of $D$'s. In such an expression, the relevant 
terms will occur at $\theta^2$ level if it involves no derivatives or 
the lowest component if it involves two derivatives. In the former case 
we will have to integrate over two $\theta$'s only. But in contrast to 
the chiral space for $N=1$ supersymmetry, there is no two-dimensional 
subspace of the $N=2$ superspace over which supersymmetry is 
represented. Hence, we are forced to consider expressions quadratic in 
$W^i_\alpha$ with two supersymmetric covariant derivatives, and to 
integrate over spacetime only. In fact, we find that the free 
Lagrangian for all the component fields is contained in the lowest 
component of
\begin{equation}
\mathscript L = - \tfrac1{192} D_i^\alpha D_{\alpha j} W^{\beta i} 
W^j_\beta + \tfrac1{192}\bar D_{\dot\alpha i} \bar D^{\dot\alpha}_j 
W^{\beta i} W^j_\beta + \text{h.c.}
\end{equation}
To be precise, we have
\begin{equation}
\mathscript L|_{\theta=\cc=0} = - \tfrac14 \mathscript F_{ab} 
\mathscript F^{ab} - \tfrac12 \partial_a \phi \partial^a \phi 
- \tfrac{1}{12} \partial_a B_{bc} \partial^a B^{bc} 
+ \tfrac12 D^2 - \tfrac14 i \lambda^{\alpha i} \ud\sigma a 
{\alpha\dot\alpha} \partial_a \bar\lambda^{\dot\alpha}_i.
\label{L1}
\end{equation}
It is straightforward to verify that Lagrangian \eqref{L1} is invariant 
under supersymmetry transformations \eqref{susy} up to a total 
divergence. Although this implies that it must be invariant under 
central charge transformation as well, since the latter is equivalent 
to two successive supersymmetry transformations, one can check directly 
that it is indeed invariant under \eqref{cc} up to a total divergence. 
This invariance of the lowest component of the Lagrangian under central 
charge and supersymmetry transformations, in fact, implies that the 
higher components are spacetime total divergences. To show this let us 
consider first the invariance under central charge transformation. 
Since the Lagrangian transforms by a total divergence under a single 
central charge transformation, it must does so under $n$ such 
transformations. That is
\begin{align}
\delta_\omega^n \mathscript L|_{\theta=\cc=0} &= \left. \left\{\omega^n 
\frac{\partial^n}{\partial\cc^n} \mathscript L \right\} 
\right|_{\theta=\cc=0} \notag \\
&= \left. \left\{\omega^n \frac{\partial^n}{\partial\cc^n} \mathscript 
L|_{\theta=0} \right\} \right|_{\cc=0} \notag \\ 
&= \omega^n \partial^a M_{n,a}(x^m).
\end{align}
Put another way, all the coefficients in the $\cc$ expansion of 
$\mathscript L|_{\theta=0}$ but the lowest one are, in fact, spacetime 
total divergences. Therefore we have
\begin{equation}
\mathscript L|_{\theta=0} = \mathscript L|_{\theta=\cc=0} + 
\sum_{n=1}^\infty \partial^a M_{n,a} \frac{\cc^n}{n!}.
\end{equation}
Although a little involved we can extend this proof to the expansion in 
$\theta$ as well. That is, it can be shown that
\begin{equation}
\mathscript L = \mathscript L|_{\theta=\cc=0} + \partial^a \mathscript 
M_a(z^M).
\end{equation}
We would like to stress that this is completely general. If the lowest 
component of a superfield is supersymmetric invariant up to a spacetime 
total divergence, all the higher components in $\cc$ as well as 
$\theta$ are spacetime total divergences. This has the immediate 
consequence that a spacetime integral of this superfield will single 
out the lowest component. In other words we can write down the 
following superfield expression for the invariant action
\begin{align}
S &= \int d^4x \left\{ - \tfrac1{192} D_i^\alpha D_{\alpha j} W^{\beta 
i} W^j_\beta + \tfrac1{192}\bar D_{\dot\alpha i} \bar D^{\dot\alpha}_j 
W^{\beta i} W^j_\beta + \text{h.c.} \right\} \notag \\
&= \int d^4x \left\{ 
- \tfrac14 \mathscript F_{ab} \mathscript F^{ab} - \tfrac12 \partial_a 
\phi \partial^a \phi 
- \tfrac{1}{12} \partial_a B_{bc} \partial^a B^{bc} 
+ \tfrac12 D^2 - \tfrac14 i \lambda^{\alpha i} \ud\sigma a 
{\alpha\dot\alpha} \partial_a \bar\lambda^{\dot\alpha}_i \right\}.
\end{align}
This represents a central charge generalization of the type of 
superactions considered in \cite{NPB-191-445}. This completes our study 
of the superfield formulation of the vector-tensor multiplet.

\section{Concluding Remarks}

We conclude by briefly mentioning the connection between the superspace
formulation of the vector-tensor multiplet and $N=1$ supersymmetry in 
six dimensions. The presence of central charge led us to augment $N=2$
superspace by two new bosonic coordinates, $z$ and $\bar z$. The 
resulting space, with six bosonic coordinates, is actually none-other 
than the superspace of $N=1$ supersymmetry in six dimensions (though 
with the fermionic coordinates treated slightly unconventionally, being 
formulated as a pair of $\SU(2)$ Majorana-Weyl spinors 
\cite{NPB-221-357,NPB-221-331,NPB-222-319}). The introduction of a 
connection $A$ in the central charge superspace then mirrors the 
formulation of supersymmetric gauge theory in six dimensions. In fact 
our first set of constraints \eqref{con-1} are precisely those used for 
the six-dimensional vector multiplet \cite{NPB-221-331,NPB-222-319}. To 
obtain the vector-tensor multiplet we imposed two further constraints. 
The first condition \eqref{4'} still has a six-dimensional 
interpretation: it corresponds to the on-shell condition for the six-
dimensional vector multiplet. It is important to note that this does 
not imply that the multiplet is on-shell in four-dimensions. Instead it 
implies a relationship between the derivatives of the component fields 
in the internal $z$--$\bar{z}$ directions and in the four-dimensional 
$x^a$
directions, as shown in equation \eqref{h2}. The second condition 
\eqref{realz} breaks the six-dimensional Lorentz symmetry. It forces 
the superfield to be independent of one of the internal $z$--$\bar{z}$ 
directions. As we have shown, together these conditions are sufficient 
to completely determine the dependence of the multiplet on the central 
charge coordinates. The connection to the component formulation of the 
vector-tensor multiplet given by Sohnius \textit{et al.} 
\cite{NPB-173-127} now also becomes clear. The second condition 
\eqref{realz} is a conventional dimensional reduction of the six-
dimensional vector multiplet down to five-dimensions. The first 
condition \eqref{4'}, corresponding to an on-shell condition in six-
dimensions, reproduces the ``dimensional reduction by Legendre 
transform'' from five to four dimensions.

A superspace formulation of the vector-tensor multiplet allows a
number of further issues to be addressed. First is the question of
duality. On shell, the usual vector multiplet and the vector-tensor
multiplet describe the same degrees of freedom. From the $N=1$ point of 
view, one is the dual of the other since the chiral multiplet in the 
usual vector multiplet is dual to the tensor multiplet in the
vector-tensor multiplet. Central charge superspace should provide a
realization of this duality in a manifestly $N=2$ supersymmetric way. A 
related question is to find an unconstrained $N=2$ prepotential for the 
vector-tensor multiplet. We also note that the form of the multiplet 
described here has been explicitly non-interacting. However, the 
superspace formalism also allows the description of non-Abelian 
multiplets as well as multiplets with gauged central charge and 
Chern-Simons terms as have been discussed in components in 
\cite{PLB-373-81}. It is also pertinent to the discussion of the most 
general form of the action for the vector-tensor multiplet. Finally the 
central charge superspace can be used to give a superfield description 
of other exotic $N=2$ multiplets, in particular the tensor and 
double-tensor multiplets mentioned in \cite{PLB-373-81}. We leave the 
discussion of these issues to future publications.

\section*{Acknowledgments}

This work was supported in part by DOE Grant No.\ DE-FG02-95ER40893 and 
NATO Grand No.\ CRG-940784.

\providecommand{\bysame}{\leavevmode\hbox to3em{\hrulefill}\thinspace}

\end{document}